\documentclass[a4 paper, 12 pt] {article}

\begin{document}

\title{\bf{Generalization of the Dick Model}}

\author{M. \'{S}lusarczyk\thanks{mslus@alphas.if.uj.edu.pl} \, and A. Wereszczy\'{n}ski\thanks{wereszcz@alphas.if.uj.edu.pl}
       \\ Institute of Physics,
        Jagellonian University, \\ Reymonta 4, Krak\'{o}w, Poland}

\maketitle

\begin{abstract}
    We discuss a model with a massless scalar coupling to the Yang-Mills SU(2) gauge field
    in four-dimensional space-time. The solutions from static, pointlike colour source are
    given. There exists not only solutions with finite energy but also singular one which
    describes confinement. The confining potential depends on the $\delta$ parameter
    of our model. The regular magnetic monopole solutions as well as the singular dyon
    configurations are also obtained. We fit the $\delta $ parameter to the experimental data.
\end{abstract}

\section{\bf{The model}}

Recently it has been pointed that in massless \cite{Dick1} and
massive \cite{Dick2} scalar field theories with a dilaton coupling
to a gauge field exist confining solutions. The potential of the
gauge field from static, pointlike colour source is in this case
singular in the spatial infinity and grows linearly as $r
\rightarrow \infty$. However, due to some phenomenological
\cite{Motyka}, \cite{Zalewski} as well as theoretical arguments
\cite{Pagels} one may also expect that the potential of
confinement is not linear. In our work we discuss a class of
models which produces relatively wide spectrum of confining
potentials.

In the present paper we would like to focus on a model described
by the action:
\begin{equation}
S = \int d^{4}x \left[ -\frac{1}{4} \left( \frac{ \phi }{\Lambda }
\right)^{8 \delta} F^{a}_{\mu \nu} F^{a \mu \nu} + \frac{1}{2}
\partial_{\mu} \phi \partial^{\mu} \phi \right],
\label{action}
\end{equation}
where $ \delta > 0$, $\Lambda $ is a dimensional constant, $\phi$
is massless scalar field coupled to the gauge field $A^{a}_{\mu}$
and $F^{a}_{\mu \nu} $ is defined in the standard manner. The
action we consider emerges from Dick model \cite{Dick2} with more
general coupling between the scalar and the gauge fields. In the
contradistinction to the Dick model our scalar field is massless
and there are not simple connections between solutions of these
models. One can add a potential or a mass term to the action but
as far as we do not know how to choose the ground state for the
scalar field it seems better and more general to consider the
model without potential. Because of the fact that we are
interested in the long range behaviour of the fields i.e. in the
small energy limit, we neglect the denominator which is present in
the original coupling. However, it is possible to consider the
full model with the denominator. Then the solutions will contain a
well-know part corresponding to the standard Yang-Mills equations.

The field equations for (\ref{action}) take the form:
\begin{equation}
 D_{\mu} \left( \frac{ \phi }{ \Lambda } \right)^{8 \delta} F^{a \mu \nu} = j^{a \nu},
\label{f_eq_1}
\end{equation}
\begin{equation}
\partial_{\mu} \partial^{\mu} \phi = -2\delta F^{a}_{\mu \nu} F^{a \mu \nu}
\frac{ \phi^{8 \delta - 1}}{\Lambda^{8\delta }}, \label{f_eq_2}
\end{equation}
where $j^{a \mu}$ is the external colour current density.
\section{The electric sector}
In this section we discuss the Coulomb solution in our model. In
the other words we find the solution of the field equations
generated by static, pointlike colour source:
\begin{equation}
 j^{a \mu} = q \delta(r) C^{a} \delta ^{\mu 0},
\label{source}
\end{equation}
where $ 1 \leq a \leq N_{c}^{2} - 1$ is an su($N_{c}$) Lie algebra
index and $C^{a}$ is the expectation value of the su($N_{c}$)
generator for a normalized spinor in colour space (see eg.
\cite{Dick1}). The field equations (\ref{f_eq_1}), (\ref{f_eq_2})
take then the form:
\begin{equation}
\left[ r^{2} \left( \frac{ \phi}{\Lambda } \right)^{8 \delta}
E^{a} \right]^{\prime} = q C^{a} \delta(r), \label{f_eq_3}
\end{equation}
\begin{equation}
\nabla^{2}_{r} \phi = -4 \delta E_{a}^{2} \frac{ \phi^{8 \delta -
1}}{\Lambda^{8\delta }}, \label{f_eq_4}
\end{equation}
where we use the standard definition $E^{ai} = - F^{a0i}$. Eqs.
(\ref{f_eq_3}), (\ref{f_eq_4}) have the following solutions
parametrized by $\beta_{0} > 0$:
\begin{equation}
\phi = A \Lambda \left( \frac{1}{\Lambda r} + \frac{1}{\beta_{0}}
\right)^{\frac{1}{1+4 \delta}}, \label{sol1_phi}
\end{equation}
\begin{equation}
E(r) = A^{-8 \delta} \frac{q}{r^{2}} \left(  \frac{1}{\Lambda r} +
\frac{1}{\beta_{0}} \right)^{- \frac{8 \delta}{1 + 4 \delta}},
\label{sol1_E}
\end{equation}
where $A = \left[ q (1 + 4 \delta ) \right]^{\frac{1}{1 + 4
\delta}}$ and $\vec{E}^{a}(r) = C^{a} E(r) \hat{r}$. \\ One can
define additional numbers for the scalar and electric field in the
standard manner (see \cite{Dick1},\cite{Cvetic}). They are given
in the following form:
\begin{equation}
Q = r^{2}E(r)|_{r \rightarrow \infty} = A^{-8 \delta} q \beta_{0}
^{\frac{8 \delta}{1 + 4 \delta}}, \label{ch_1}
\end{equation}
\begin{equation}
D = -r^{2}\frac{d \phi}{dr}|_{r \rightarrow \infty} =
\frac{A}{1+4\delta} \beta_{0}^{\frac{4 \delta}{1 + 4 \delta}}.
\label{ch_2}
\end{equation}
It is possible to regard the numbers as some effective charges of
the scalar and electric field. These numbers are not independent
and they satisfy the following condition:
\begin{equation}
\frac{D^2}{Q} = q \label{cond}
\end{equation}
The values of $D$ and $Q$ are finite for each, finite $\beta_{0}$.
The existence of the effective charges is rather mysterious but it
can be understood from symmetries point of view.  Namely, using
equation (\ref{f_eq_3}) we can eliminate the electric field in
(\ref{f_eq_4}). It is easy to notice that after rewriting this
equation in terms of the variable $x=\frac{1}{r} $ a new symmetry
appears. This symmetry is the translational symmetry of the
variable $x$,
\begin{equation}
x \longrightarrow x'=x+x_0 \, \, \, \, \phi(x) \longrightarrow
\phi'(x')=\phi(x). \label{sym}
\end{equation}
The generator of the symmetry has form: $\hat{D}_x =
-\frac{d}{dx}$, or, in the old variable $r$:
\begin{equation}
\hat{D}_r = r^2 \frac{d}{dr} \label{gensym}
\end{equation}
In the natural way we define the effective scalar charge using the
generator (\ref{gensym}), $D=-\hat{D}_r \phi(r) \mid_{r
\rightarrow \infty } $ which means that the scalar charge emerges
from the symmetry (\ref{sym}). It is possible to break the
symmetry by adding same new terms to the action. In the simplest
case one can introduce a potential for the scalar field. Moreover,
this potential fixes asymptotic behaviour of the scalar field i.e.
the value of the parameter $\beta_{0}$ as well as the effective
scalar charge. However, in our work we do not break the symmetry
and continuous spectrum of the solution survives.

The energy density for the solutions (\ref{sol1_phi}),
(\ref{sol1_E}) takes the form:
\begin{equation}
\varepsilon = A^{-8 \delta} \frac{q^{2}}{r^{4}} \left(
\frac{1}{\Lambda r} + \frac{1}{\beta_{0}} \right)^{- \frac{8
\delta}{1 + 4 \delta}}. \label{energy_dens}
\end{equation}
Thus we may conclude that the energy of configurations
(\ref{sol1_phi}) - (\ref{sol1_E}) is finite for $ 0 < \beta_{0} <
\infty$ only if the parameter $ \delta > \frac{1}{4}$. In this
case one finds:
\begin{equation}
E_N = \int \varepsilon r^{2} dr = \Lambda \frac{4 \delta + 1}{4
\delta - 1} A^{-8 \delta} q^{2} \beta_{0}^{\frac{4 \delta - 1}{4
\delta + 1}}. \label{energy}
\end{equation}
From the full spectrum of Coulomb solutions the family with finite
energy and charges (\ref{ch_1}), (\ref{ch_2}) can be then
separated. As the energy depends on the parameter $\beta_0$ it can
achieve any positive value. On the other hand there exists the
singular solution:
\begin{equation}
\phi(r) = A \Lambda  \left( \frac{1}{\Lambda r}
\right)^{\frac{1}{1 + 4 \delta}}, \label{sol2_phi}
\end{equation}
\begin{equation}
E(r) = A^{-8 \delta} q \Lambda^2 \left( \frac{1}{\Lambda r}
\right)^{\frac{2}{1 + 4 \delta}}, \label{sol2_E}
\end{equation}
which describes the confining sector of our theory. The
colour-electric potential has the following form:
\begin{equation}
 V(r) = \left\{
\begin{array}{cc}
   \frac{4\delta +1}{4\delta -1}
  A^{-8\delta } q \Lambda^{\frac{4\delta }{4\delta +1}} \cdot
  \, r^{\frac{4\delta -1}{4\delta +1}} & \delta > \frac{1}{4} \\
  \\
  \Lambda A^{-8\delta } q \ln \Lambda r & \delta =\frac{1}{4}
\end{array}
\right. \label{potential}
\end{equation}
The asymptotic behaviour of $V(r)$ changes from $\log r$ for
$\delta = \frac{1}{4}$  to the linear potential in the limit
$\delta \rightarrow \infty$. Obviously, the limit $\delta
\rightarrow \infty$ cannot be implemented on the Lagrangian level.
It is not possible to realize the linear flux-tube but it may be
approximated with arbitrary accuracy taking sufficiently large
value of $\delta$. The energy density for the singular solution
may be obtained from (1\ref{energy_dens}) in the limit $\beta_{0}
\rightarrow \infty$. It should be stressed that on the contrary to
the standard Coulomb potential the singularity of energy and
charges is caused by the behaviour in the spatial infinity. In
fact the energy density is singular at $r=0$ but this singularity
is integrable for $\delta > \frac{1}{4}$.

Elimination of single charge states from the physical spectrum
theory is not sufficient to  have confinement -- like sector in
our model. One have to check that a state with two, opposite
colour charges, has a finite energy. In that case we shall
consider a dipole-like external charge,
\begin{equation}
j^{a \mu }= q[\delta (z+\frac{a}{2} ) -\delta (z-\frac{a}{2} )]
\delta (x) \delta (y) \delta^{a3} \delta^{\mu 0}. \label{dipol}
\end{equation}
We restrict consideration to Abelian sector of the theory. Because
of nonlinearity of the equations of motion we are not able to
solve them in analytically. However, as it was presented in
\cite{Adler1}, \cite{Adler2}, \cite{Arodz}, some numerical methods
can be applied.

It is convenient to define the dual potential $\vec{C} $:
\begin{equation}
\vec{D}=\nabla \times \vec{C}, \label{vector}
\end{equation}
where $\vec{D}=(\frac{\phi}{\Lambda })^{8\delta } \vec{E} $ is the
dielectric induction. After introduction of the cylindrical
coordinates $(\rho, \alpha, z)$ we assume that the dual potential
has the form:
\begin{equation}
\vec{C}=\frac{\hat{\alpha}}{2\pi \rho} \Phi(\rho ,z),
\label{scalar}
\end{equation}
where $\hat{\alpha}$ is the unit vector tangent to the $\alpha$
coordinate line and $\Phi$ is a scalar flux function. Then the
equation of motion can be rewritten as:
\begin{equation}
\nabla \left( \frac{1}{\rho } \left(\frac{\phi }{\Lambda}
\right)^{-8\delta } \nabla \Phi \right)=0, \label{eqnew}
\end{equation}
\begin{equation}
\nabla^2 \phi +4\delta \left( \frac{\phi }{\Lambda }
\right)^{-8\delta } \frac{\phi }{\rho } \mid \nabla \Phi \mid^2=0.
\label{eqnew2}
\end{equation}
Following \cite{Adler1}, \cite{Adler2} we fix boundary conditions
as:
\begin{equation}
\begin{array}{cc}
  \Phi=0 & \rho=0,\,\, \mid z \mid > R/2 \\
  \Phi=q & \rho=0,\,\, \mid z \mid < R/2.
\end{array}
\label{boundary}
\end{equation}
Moreover, we put that $\Phi \rightarrow 0, \phi \rightarrow 0$ for
$\rho^2 +z^2 \rightarrow \infty$. This set of equations can be
solved numerically. Fig. 1 presents the flux function $\Phi$
computed for $900 \times 900$ mesh, $\delta = 0.75$ and $q =1.1$
(for detailed description of the applied numerical procedure see
 \cite{Adler2}, \cite{Arodz}).

It is possible to construct examples of the flux  and the scalar
function which obey the boundary condition and have finite energy:
\begin{equation}
\Phi=\frac{q}{2} \left( \frac{z+R/2}{\sqrt{\rho^2 +(z+R/2)^2}}
-\frac{z-R/2}{ \sqrt{\rho^2+(z-R/2)^2}} \right),
\label{minenergy1}
\end{equation}
\begin{equation}
\phi=A\Lambda \left( \frac{1}{\Lambda \sqrt{\rho^2 +(z+R/2)^2}}
\right)^{ \frac{1}{1+4\delta}} -A\Lambda \left( \frac{1}{\Lambda
\sqrt{\rho^2 -(z-R/2)^2}} \right)^{\frac{1}{1+4\delta}}.
\end{equation}
As it was mentioned in \cite{Arodz}, in spite the fact that these
functions do not obey Eqs. (\ref{eqnew}), (\ref{eqnew2}) they give
an upper bound for the total field energy for the charges $q,-q$.
One can check that
\begin{equation}
E_{N} \sim R^{\frac{4 \delta - 1}{4 \delta  + 1}}.
\end{equation}
\section{The magnetic sector}

Let us now consider purely magnetic, non-abelian content of our model.
We use typical, spherically-symetric Ansatz:
\begin{equation}
A^{a}_{i} = \epsilon_{aik} \frac{x^{k}}{r^{2}}(g - 1),
\;\;\;\;A^{a}_{0} = 0, \label{ansatz1}
\end{equation}
where $g(r)$ is a function of radial coordinate only. Inserting
the Ansatz (\ref{ansatz1}) into Eqs. (\ref{f_eq_1}),
(\ref{f_eq_2}) we get:

\begin{equation}
\left[ \phi^{8 \delta} g' \right]' + \frac{\phi^{8 \delta}}{r^{2}}
g \left( 1 - g^{2} \right) = 0, \label{f_eq_5}
\end{equation}
\begin{equation}
-\frac{1}{r^{2}} \left( r^{2} \phi' \right)' + 4 \delta \frac{
\phi^{8 \delta - 1}}{\Lambda^{8\delta }} \left[
\frac{2g'^{2}}{r^{2}} + \frac{(g^{2} - 1)^{2}}{r^{4}} \right] = 0.
\label{f_eq_6}
\end{equation}
The vacuum state corresponds to $\phi = 0$ and arbitrary value of
the gauge field or to $g = \pm 1$ and $\phi = \phi_{0}$, where
$\phi_{0}$ is an arbitrary constant. The standard magnetic
monopole solution may be constructed by taking $g = 0$. Then the
scalar field is given by a family of configurations parametrized
by $\beta_{0}$:
\begin{equation}
\phi (r) = B \Lambda \left(  \frac{1}{\Lambda r} +
\frac{1}{\beta_{0}} \right)^{\frac{1}{1 - 4 \delta}},
\label{sol3_phi}
\end{equation}
where $B = (1 - 4 \delta )^{\frac{1}{1 - 4 \delta}}$. The energy
density reads:
\begin{equation}
\varepsilon = \frac{B^{2}}{(1 - 4 \delta ) ^{2}} \frac{1}{r^{4}}
\left(  \frac{1}{\Lambda r}
 + \frac{1}{\beta_{0}} \right)^{\frac{8 \delta}{1 - 4 \delta}},
\label{ene_dens_2}
\end{equation}
It is easy to check that for each $ 0 < \beta_{0} < \infty$ the
energy for such a configuration is finite.
\begin{equation}
E_N = \Lambda \frac{4 \delta - 1}{4 \delta + 1} B^{8 \delta} q^{2}
\beta_{0}^{\frac{4 \delta + 1}{4 \delta - 1}}. \label{energymag}
\end{equation}
The effective scalar charge takes the value:
\begin{equation}
D = \frac{B}{1-4 \delta} \beta_{0}^{\frac{-4 \delta}{1 - 4
\delta}}.
\end{equation}
Of course there exist also the infinity energy solution, namely:
\begin{equation}\label{magsing1}
  \phi (r)=B \Lambda \left( \frac{1}{\Lambda r} \right)^{\frac{1}{1-4\delta }}
\end{equation}
In the magnetic case, for $\delta \geq \frac{1}{4}$,  the
asymptotic divergence of the singular solution is much bigger than
in the electric case. Because of that magnetic singular monopole
can not form finite energy bound states. The electric Coulomb
sector as well as the magnetic one are related to each other by
the duality transformation \cite{Cvetic}:
\begin{equation}
F^{a}_{\mu \nu} \rightarrow \phi^{8 \delta} F^{*a}_{\mu \nu},
\;\;\; \delta \rightarrow - \delta
\end{equation}
\section{The dyon}
Let us now consider generalization of Ansatz (\ref{ansatz1}) with
electric field:
\begin{equation}
A^{a}_{0} = \frac{x^{a}}{r} h \label{ansatz2}
\end{equation}
The field equations take then the form:
\begin{equation}
  \label{d2}
   [\phi^{8\delta } g']' =\frac{\phi^{8 \delta }}{r^2} [g(g^2 -1) +r^2 h^2 g]
\end{equation}
\begin{equation}
  \label{d3}
  [r^2 \phi^{8 \delta } h']' =2\phi^{8 \delta } h g^2
\end{equation}
\begin{equation}
   -\frac{1}{r^2} (r^2 \phi' )' +4\delta \frac{ \phi^{8\delta  -1}}{\Lambda^{8\delta }}
    \left[ \frac{2g'^2
  }{r^2} + \frac{(g^2-1)^2}{r^4} -2 \frac{h^2 g^2 }{r^2} -h'^2 \right] =0
\label{f_eq_7}
\end{equation}
The dyon solution is given by the formulae:
\begin{equation}
g = 0,  h' = C \left( \frac{\phi }{\Lambda } \right)^{-8 \delta}
\frac{1}{r^{2}}, \label{sol_dyon}
\end{equation}
where G is an arbitrary constant and $\phi$ satisfies the
equation:
\begin{equation}
- \frac{1}{r^{2}} \left( r^{2} \phi' \right)' + 4 \delta \frac{
\phi^{8 \delta - 1}}{\Lambda^{8\delta }} \left[ \frac{1}{r^{4}} -
G^{2} \left( \frac{\phi}{\Lambda } \right)^{-16 \delta}
\frac{1}{r^{4}} \right] = 0. \label{eq_dyon}
\end{equation}
The dyon solution can be interpreted as standard magnetic monopole
surrounded by the region of colour electric field. From Eq.
(\ref{eq_dyon}) one easily gets:
\begin{equation}
\pm \frac{d \phi}{d x} = \phi^{4 \delta} - G \phi^{-4 \delta}.
\label{eq_dyon_1}
\end{equation}
There exist only a few values of parameter $\delta$ for which Eq.
(\ref{eq_dyon_1}) can by solved analytically. Otherwise only
numerical estimations are known.For example choosing $\delta =
\frac{1}{4}$ the solution reads:
\begin{equation}
\phi (r) = \Lambda \sqrt{G + (g_{0} - 1)e^{\mp \frac{2}{\Lambda
r}}}.
\end{equation}
Here $g_{0}$ is a constant and the charges are
\begin{equation}
  \label{d81}
  D=\frac{-g_0}{\sqrt{G+g_0}}, \; \; \; Q=\frac{G}{G+g_0}.
\end{equation}
Analogously for $\delta = \frac{1}{2}$ one gets:
\begin{equation}
\arctan \frac{\phi }{\Lambda } - \mbox{arcth} \frac{\phi }
{\Lambda} = 2 \left(  \frac{1}{\Lambda r} + \frac{1}{\beta_{0}}
\right).
\end{equation}
It is easy to show that energy for all dyon solutions is infinite.
Unfortunately, singularity of the energy emerges for $r=0$ and
there is no chance for two-dyons, finite energy solutions.
\section{Conclusions}

The main result of our work is that the electric singular
solutions can form the dipole-like finite energy states i.e.
something like a confining sector of the model emerges: the static
solutions with non-vanishing total charge are excluded from the
physical spectrum while the $q,-q$ dipoles can appear. For the
discussed model one can find, in the Coulomb sector of the theory,
the potential of confinement (\ref{potential}). The  behaviour of
the energy depends on the model parameter $\delta$. Taking into
account the form of confining potential which is suggested by
phenomenological arguments (see eg. \cite{Motyka}) the parameter
$\delta$ can be fixed. For $V(r) \sim \sqrt{r}$ one finds $\delta
= \frac{3}{4}$. On the other hand, there is continuous spectrum of
finite energy solutions. Because of the fact that these solutions
describe a source with a fixed charge it can be possible to look
at them as screening (see eg. \cite{kiskis}, \cite{sikivie}).

For the magnetic, sourceless part of the model, it is also
possible to separate the finite as well as the infinite energy
sector. One can interpret spherically-symmetric, sourceless
solutions with finite energy, which form the finite energy sector,
as classical glueballs \cite{Kerner}. On the contrary to electric
part the magnetic singular monopoles can not form finite energy
bound states. So in the sense, magnetic part of the theory does
not provide the confinement.

The confinement sector has its equivalent in the Pagels-Toumbulis
model \cite{Pagels}, \cite{Arodz}. Our scalar field plays the same
role as the dielectric function in the effective action
\cite{Pagels}. Of course both models are not identical. For
example there is not (probably) any screening sector in
Pagels-Toumbulis model.

 As we see the model (\ref{action}) posses two phases: confinement and
screening (or glueball)-like. The phases refer to different,
asymptotic values of the scalar field, zero or non-zero
respectively. The asymptotic value of the scalar field can be
fixed by a potential term. Then the confinement phase could be
understood as the symmetric phase whereas the screening sector as
the phase with broken symmetry. Unfortunately, we do not know the
form of the potential. However, color dielectric models
\cite{coldiel} can give some hints \cite{My}.

We would like to thank Professor H. Arod\'{z} for reading the
manuscript and for many stimulating discussions.

\begin{figure}[ht]
    \unitlength 1cm
        \begin{picture}(0,10)
        \put(-3.0,-0.6){
            \includegraphics{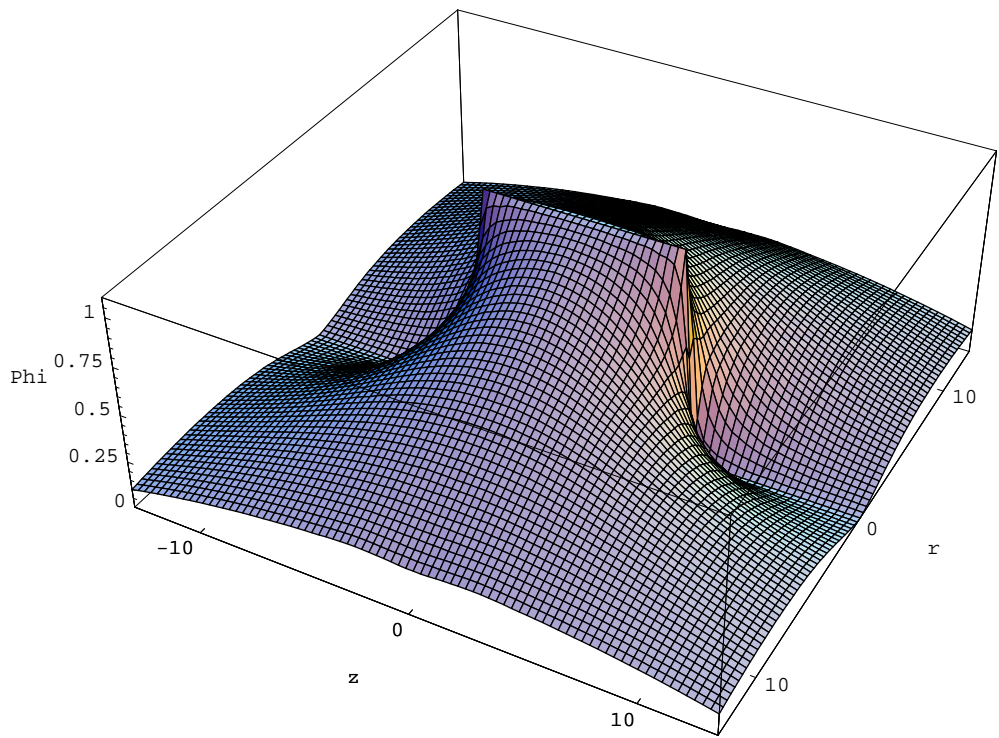}           }
    \end{picture}
    \caption{The flux function $\Phi$ for $\delta = 0.75$ and $R = 10$}
\end{figure}


\begin{thebibliography}{19}

\bibitem{Dick1} R.Dick, Phys. Lett. {\bf B397}, (1996), 193; R.Dick Phys. Lett {\bf B409}, (1997), 321.
\bibitem{Dick2} R. Dick Eur. Phys. J {\bf C6}, (1999), 701; M. Chabab, R. Markazi, E. H. Saidi,
hep-th/0003225.
\bibitem{Motyka} L. Motyka, K. Zalewski, Z. Phys. {\bf C69}, (1996), 343.
\bibitem{Zalewski} K. Zalewski, Acta Phys. Pol. {\bf B29}, (1998), 2535.
\bibitem{Pagels} H. Pagels, E. Tomboulis, Nucl. Phys. {\bf B43}, (1978), 485.
\bibitem{Adler1} S. L. Adler, T. Piran, Rev. Mod. Phys. {\bf 56}, (1985), 1.
\bibitem{Adler2} S. L. Adler, Phys. Rev. {\bf D20}, (1981) 3273.
\bibitem{Adler3} S. L. Adler, T. Piran, Phys. Lett. {\bf B113}, (1982), 405.
\bibitem{Arodz} H. Arod\'{z}, M. \'{S}lusarczyk, A.Wereszczy\'{n}ski,
 Acta Phys. Pol.{\bf B32}, (2001), 2155
\bibitem{Cvetic} M. Cvetic, A. A. Tseytlin, Nucl. Phys. {\bf B416},
(1994), 137.
\bibitem{kiskis} J. Kiskis, Phys. Rev. {\bf D21}, (1980), 421.
\bibitem{sikivie} P. Sikive, N. Weiss, Phys. Rev. Lett. {\bf 40}, (1978), 1411;
P. Sikive, N. Weiss, Phys. Rev. {\bf D18}, (1978), 3809.
\bibitem{Kerner} D. Gal'tsov, R. Kerner, Phys. Rev. Lett. {\bf 84}, (2000), 5955.
\bibitem{coldiel} H. Arod\'{z}, H. J. Pirner, Acta Phys. Pol.
{\bf B30}, (1999), 3895.
\bibitem{My} M. \'{S}lusaraczyk, A. Wereszczy\'{n}ski, work in progress.
\end{thebibliography}
\end{document}